# 14
# AI AS IA

## The use and abuse of artificial intelligence (AI) for human enhancement through intellectual augmentation (IA)

*Alexandre Erler and Vincent C. Müller*

## 14.1 Introduction

### *14.1.1 Human enhancement and intellectual augmentation*

It is now widely agreed that we live in the age of artificial intelligence (AI). This paper will discuss the potential and the risks of using AI to achieve human enhancement and what we shall call intellectual augmentation (IA). Let us begin with some clarifications on how we propose to understand those key concepts. Broadly in line with previous work of ours (Erler and Müller, forthcoming), we shall define human enhancement as encompassing technological interventions that either:

  a) Improve aspects of someone's functioning beyond what is considered "normal," or
  b) Give that person new capabilities that "normal," non-enhanced humans do not possess.

Our proposed definition is compatible with the so-called therapy-enhancement distinction, insofar as it denies that "pure" therapies (i.e. interventions that restore or maintain health without meeting either condition 1 or 2) count as enhancements, even though they do improve human functioning in some way. That said, we also believe that the therapy-enhancement distinction should not be understood as entailing a strict dichotomy between these two types of interventions. The existence of a hybrid category of "therapeutic enhancements" should also be acknowledged (Wolbring et al., 2013; Erler and Müller, forthcoming), and we shall see that it includes some applications of AI for HE. Such hybrid interventions either restore or preserve normal functioning in a manner that matches either condition 1 or 2 above. Consider, for instance, the vaccines against COVID-19, which are meant to protect our health by endowing us with a capacity – an immunity to infection against that virus – that is not part of the "normal" human condition.

    A distinction is often drawn between enhancements in the full sense and mere useful "tools" (Lin et al., 2013; Erler and Müller, forthcoming). The former, but not the latter, are assumed to help produce desired outcomes by truly altering a person's physical or cognitive functioning. It might thus be said that a tool like a calculator, while helping us reach the correct result when performing a complex multiplication, nevertheless does not do so by improving our mathematical abilities or general cognitive functioning. Rather, the calculator relieves us of the need to engage in mathematical reasoning by performing that task for us. (That said, we will consider a possible challenge

    

to this view when discussing the extended mind thesis in Section 14.3.4) As we will see, one might suspect that nearest-term applications of AI aimed at improving human decision-making beyond the "normal" will not count as true enhancements, as they will fail to meet the requirement of altered functioning. To take this possibility into account, we propose to recognize a broader category of improvements in decision-making and mental performance that includes enhancements: what we term "intellectual augmentation." Forms of IA that meet the altered functioning requirement also count as enhancements in our view, whereas those that do not are "simply" IA.

Applications of AI for enhancement can either have *broad* or *narrow* effects: they might for instance improve broad human capacities like memory or general intelligence, or they might target narrower aspects of cognitive functioning – say, clinical judgment. Finally, our proposed understanding of IA can be described as "liberal" insofar as the interventions it includes can produce their desired effects either *directly* or *indirectly*. For example, a brain-computer interface (BCI) that directly boosted a person's ability to focus by applying some form of brain stimulation, and a different BCI that indirectly improved that capacity by simply alerting the subject to when she got distracted, will both count as IA ("attention augmentation") on our approach.

### 14.1.2 Defining AI

The term "artificial intelligence" (AI) is now used in two main meanings:

(a) AI is a research program to create computer-based agents that can show complex behavior, capable of reaching goals (McCarthy et al., 1955), and
(b) AI is a set of methods employed in the AI research program for perception, modeling, planning, and action: machine learning (supervised, reinforced, unsupervised), search, logic programming, probabilistic reasoning, expert systems, optimization, control engineering in robotics, neuromorphic engineering, etc. Many of these methods are also employed outside the AI research programme (Russell, 2019, Russell and Norvig, 2020, Görz et al., 2020, Pearl and Mackenzie, 2018).

The original research program (a) from the "Dartmouth Conference" in 1956 onwards was closely connected to the idea that computational models can be developed for the cognitive science of natural intelligence and then implemented on different hardware, i.e. on computing machines. This program ran into various problems in the "AI Winter" ca. 1975–1995, and the word "AI" got a bad reputation; it thus branched into several technical programs that used their own names (pattern recognition, data mining, decision support system, data analytics, cognitive systems, etc.). After 2000, AI saw a resurgence, with faster hardware, more data, and an emphasis on neural network machine-learning systems. From ca. 2010, "AI" became a buzzword that resonated in circles outside computer science; now everyone wants to be associated with AI. As a result, the meaning of "AI" is currently broadening toward (b). It is this second, broader sense of AI that we will be relying on in our discussion. We shall consider various systems and devices, both present and foreseen, capable of complex information processing and targeted at the pursuit of certain human goals (or the maximization of expected utility).

Some view the ultimate endpoint of the AI research program as the attainment of machine "superintelligence." Superintelligence is typically explained on the basis of *general human intelligence*, where "super" intelligence is just *more of the same:*

> We can tentatively define a superintelligence as any intellect that greatly exceeds the cognitive performance of humans in virtually all domains of interest ... Note that the definition is noncommittal about how the superintelligence is implemented. It is also noncommittal regarding qualia; whether a superintelligence would have subjective con-scious experience might matter greatly for some questions (in particular for some moral questions), but our primary focus here is on the causal antecedents and consequences of superintelligence, not on the metaphysics of mind.
   (Bostrom, 2014).

Besides superintelligence strictly understood, which entails a form of artificial general intelligence (AGI), Bostrom also envisages the possibility of "domain-specific" superintelligences (*ibid*.) – AI systems that vastly surpass human performance in specific cognitive domains yet cannot be applied outside of that narrow scope. Deep Blue, the chess-playing supercomputer that beat Gary Kasparov in 1997, and AlphaZero, the software developed by Google to achieve superhuman performance at the games of Go, chess, and shogi, are contemporary examples of domain-specific or narrow superintelligence (although Deep Blue is even narrower than AlphaZero). While we will mostly eschew highly speculative scenarios involving the rise of full-fledged superintelligence, which does not appear to be a likely near-term development, we will occasionally consider the prospect of narrower forms of it, as we will see in the next section, outlining the main possible applications of AI for IA.

## 14.2  AI technologies of relevance to the prospect of IA

### 14.2.1  AI advisors/"outsourcing"

The first relevant application is the use of "AI advisors," or what some have called AI outsourcing (Danaher, 2018), to improve human decision-making. AI advisors can be viewed as the logical next step from familiar tools like GPS or the internet. Their existing instantiations include, among others:

a) Well-known virtual assistants like Apple's Siri or Amazon's Alexa, which can help users decide, in line with their personal preferences, which product to buy, which restaurant to go to in an unfamiliar area, etc.
b) Clinical decision support systems, designed to assist healthcare professionals with clinical decision-making and increase the probability of reaching an accurate diagnosis (Sutton et al., 2020)
c) "Robo-advisors," which provide automated, algorithm-based assistance with financial planning, often at a lower cost than traditional human advisors (Frankenfield, 2021).

The kinds of decision-making ability that existing AI assistants or advisors aim to improve tend to be quite narrow, which of course does not refute their importance. Yet as such systems become more sophisticated with further technological development, their range of application might widen. The capacity for *moral* decision-making has received special attention in the academic literature. Some authors have thus proposed to create an AI system that would constantly monitor an agent's physiology, mental states, and environment, and on that basis, make her aware of potential biases in her decision-making or suggest the best course of action in a given situation (Savulescu and Maslen, 2015). To provide such assistance, the moral AI advisor would rely on the agent's own stated values, the implications of which it would draw using its superhuman capacity for information processing (*ibid*.; Giubilini and Savulescu, 2018). Other authors advocate for an alternative approach, involving a moral AI that acted like a Socratic interlocutor, spurring the user to think more carefully and thoroughly (Lara and Deckers, 2020).

The concept of a moral AI advisor does not intrinsically entail the monitoring of the advisee's brain states, yet the proposals just outlined all involve such monitoring, which would likely require the sort of technology discussed in the next section. Furthermore, while some aspects of these proposals do not presuppose the development of superintelligent AI systems (e.g. alerting the user that her level of tiredness might impair her moral reasoning), others do seem to do so. For instance, Giubilini and Savulescu describe their proposed moral AI advisor as "an expert more informed and more capable of information processing than any other human moral expert we trust" (2018, p. 177). While such a system need not represent a form of AGI, it would nevertheless demonstrate superhuman performance with regard specifically to moral reasoning.

Such a development would arguably have momentous practical implications. At the very least, if technology advances to the point where creating superintelligent *moral* AI advisors becomes possible, then we may expect the same to be true of other superintelligent AI advisors that can assist with other aspects of human decision-making: those not belonging to morality strictly understood, but rather to what philosophers would call *prudence*. As evidenced by the popularity of self-help books, coaching, and financial advisors, people do not simply seek to become more moral but also professionally, financially, and socially successful, happier, fitter, and healthier. It is not clear that any of these pursuits call for more complex reasoning capabilities than those required for the provision of sound moral advice.

Should we then anticipate that people will consult different AI advisors targeted at different goals, as they might currently take advice from "experts" in various domains, and then balance input from different sources (perhaps asking their moral AI advisor to help with such balancing)? This will partly depend on whether such stable co-existence of (relatively narrow) superintelligent AI systems is at all possible. Some might conjecture that their development would soon be followed by the arrival of AGI, raising, in turn, the prospect of an "intelligence explosion" (Good, 1965) with radically transformative consequences for the world – what is often labeled the "singularity" (Kurzweil, 2005). Since it is hard to foresee what such an outcome would look like, we will not discuss it further in what follows, yet it is worth noting that its possibility cannot be ruled out.

Even a scenario including only domain-specific superintelligent AIs could cause significant social disruption, threatening the relevance of many human occupations (including, to some extent, professional ethicists!). One response to this challenge might be to try and foster greater integration between humans and AI, using for instance a technology like brain-computer interfaces (BCIs).

### *14.2.2 Brain-computer interfaces*

BCIs are designed to establish a direct connection between a person's brain and a computer. *Invasive* BCIs do so using implanted electrodes that require surgery. *Non-invasive* variants, by contrast, capture brain activity using techniques like electroencephalography (EEG), functional Magneto-Resonance Imaging (fMRI), or magnetoencephalography (MEG). While invasive BCIs allow for higher resolution readings, they also tend to be riskier, as we will see later (Ramadan et al., 2015). BCIs are basically used to record and interpret brain activity. In the case of "bi-directional" BCIs, which at the time of writing are still at the research and development stage, they can additionally be used to modulate brain activity via electric stimulation (Hughes et al., 2020).

BCIs are relevant to our discussion through their association with AI, for instance since AI algorithms are used to interpret the brain data they collect. They are primarily designed for therapeutic purposes, and their potential in that regard seems very broad. As demonstrated by various studies, they thus offer promise to people with paralysis, allowing them for instance to control tablet devices (Nuyujukian et al., 2018), wheelchairs, speech synthesizers (Cinel et al., 2019), and robotic or prosthetic limbs (Vilela and Hochberg, 2020), as well as powered exoskeletons (He et al., 2018), using their thoughts. Other potential beneficiaries of BCI technology include patients with residuals of stroke, traumatic brain injury, spinal cord injury, and memory disorders (DARPA, 2018; Klein, 2020).

Uses of BCIs for IA are already a reality, however. Some of the uses just described arguably constitute therapeutic enhancements. While the ultimate purpose behind the BCIs used by para-lyzed patients may be therapeutic in nature (restoration of the "normal" ability to use a computer), it is nevertheless achieved via the conferral of a "supernormal" capacity – control of an external device via thought. Existing applications of BCIs for *pure* IA include the use of EEG-based brain-monitoring BCIs to improve capacities like attention and emotional regulation. In the educational context, some thus expect that monitoring the level of attention and engagement of students can help adjust the learning process, and ultimately optimize these different factors (Williamson, 2019). One company called BrainCo thus designed an EEG headband aimed at providing precisely such data to teachers. The company claims the data can then be used to design focus-training games that will enhance users' capacity for sustained attention (BrainCo, 2020) – although the validity of such claims has been questioned. The headband has already been trialed in some Chinese schools, causing controversy (Jing and Soo, 2019).

Looking into the future, there is a broad interest, among various sectors, in using BCIs to monitor the brain activity of employees, for the ultimate purpose of improving performance and productivity. It has thus been suggested that such systems could be designed to automatically adjust environmental conditions such as room temperature, to maximize a worker's efficiency (Valeriani et al., 2019). Another major context in which BCIs are likely to be used for IA is the military. Both the US and Chinese military are reportedly considering the potential of BCIs to enable the direct transfer of thoughts from brain to brain, thereby the ability to communicate silently, as well as allowing for faster communication and decision-making among soldiers and military commanders (Kania, 2019). Other relevant applications include the direct control of semiautonomous systems and drone swarms via thought, and more ambitiously, disruption of pain and regulation of emotions like fear among warfighters – although the latter two applications would require BCI systems featuring some form of brain stimulation (Binnendijk et al., 2020).

On an even more futuristic note, entrepreneurs like Elon Musk have made the headlines by proclaiming their ambition to develop BCIs that would enhance human cognition to the point of ultimately yielding "superhuman intelligence" (Lewis and Stix, 2019). Musk's reasoning is the standard one we have outlined already: continuous progress in AI ultimately threatens to render humans obsolete unless they choose to radically augment themselves by merging with AI. The details of how this process of radical augmentation is to take place are, however, much less clear, at least for now.

Having laid out some of the main foreseeable applications of AI for IA, we now turn to a (necessarily brief) overview of their potential ethical ramifications.

## 14.3 Philosophical and ethical issues pertaining to the use of AI for IA

### *14.3.1 Devices not performing as expected*

Some have argued that the most relevant present-day concern about consumer EEG headsets is that companies selling them tend to misrepresent their enhancing effects (McCall and Wexler, 2020). To some extent, this concern also applies to future AI advisors: the quality of the advice provided by such a system could in principle fail to meet the standards promised by its manufacturer. This could happen either because of inadequate design, over-hyping of existing technical capabilities, or because the company selling the device had purposefully programmed it to nudge users toward courses of action that were favorable to its own interests, or those of its business partners, even when these were not fully in line with a user's ethical commitments (Bauer and Dubljevic, 2020). Moreover, besides such purposeful distortions of results, biases could also be inadvertently introduced into the underlying algorithms – we consider this latter issue in Section 14.3.5.

In addition to users exercising critical judgment, supplemented with feedback from acquaintances and reviews from other users as well as experts in the relevant domain, forms of quality con-trol that would help mitigate those concerns might include a (presumably optional) certification process, analogous to that applicable to the human equivalents of AI advisors. We are already seeing this idea being implemented in some areas, such as the financial sector in Norway (Iversen, 2020). While moral AI advisors might present unique challenges in this context, the existence in today's world of standardized tests of critical thinking ability (Hitchcock, 2018), and university and high school courses in ethics, suggests that these challenges need not be intractable. Perhaps the "pool of moral experts" whom Giubilini and Savulescu suggest could be consulted when programming such devices (2018, p. 177) could also help design an appropriate certification procedure. Finally, ensuring *transparency* in the functioning of those devices would be key to quality control: they should be designed so as to always provide the user with a detailed justification for any particular recommendation they might offer. Nevertheless, this goal might present challenges related to the design of AI systems (O'Neill et al., forthcoming).

Concerns about efficacy might be especially salient with regard to BCIs, particularly in contexts like those outlined above, in which people might face coercive pressures to use them. Taking for instance the EEG headsets currently used in certain places to monitor the attention levels of students and workers, the reliability with which such devices can measure brain activity has been questioned (McCall and Wexler, 2020). They might conceivably have positive effects on attention even if their measurements are not accurate, simply because of the users' awareness that they are being monitored, combined with the expectations of others. Yet this would still not mean that they were truly efficacious. Furthermore, some worry that even when BCIs can reliably measure attention levels, they might still not – at least for now – allow us to determine whether a user is actually focused on their work or study, as opposed to, say, on their mobile phone (Gonfalonieri, 2020). Since it would clearly be problematic if rewards and punishments were to be meted out, e.g. by employers, based on misleading brain readings, promoting or even enforcing adequate quality standards might be especially important in such contexts.

### 14.3.2  *Safety, coercion, and responsibility*

Concerns about safety are mostly relevant to BCIs. First, invasive BCIs present health risks such as scarring, hemorrhaging, infection, and brain damage (Ramadan et al., 2015). Bi-directional BCIs involving electrical brain stimulation might also raise safety concerns, although this would depend on their *modus operandi*. The existing scientific literature suggests that less invasive forms of stimulation such as transcranial direct current stimulation (tDCS) have a good safety profile among both healthy and neuropsychiatric subjects, although uncertainties remain regarding long-term use and increased exposure (Nikolin et al., 2018). Secondly, the malfunction of a BCI, or its hijacking by a malicious third party, could result in harm to the user or to other people.

Addressing the first issue will likely require enforcing standards of good practice for pure enhancement uses of invasive BCIs similar to those already governing their therapeutic applications. These would include appropriate licensing requirements for those performing the needed surgical procedures, as well as a certification process for the implanted devices. Dealing with the second issue requires establishing regulations to promote adequate security standards in the design process, and to hold BCI manufacturers liable when malfunctioning devices have harmful consequences in cases where it is clear that the user herself bears no responsibility for what happened. Admittedly, this still leaves us with cases of a trickier kind: for instance, even if we could determine, based on a BCI recording, that the harmful command had its ultimate source in the brain of the user, would that automatically mean that we could hold that person fully responsible? Could such a command ever be triggered by an "automatic" thought over which they would exert only limited or no control (Burwell et al., 2017)? This, in turn, raises the question of whether such concerns can be alleviated simply via proper BCI design, or whether more will be needed (e.g. new legal provisions).

The issue of safety arguably becomes trickier when it is coupled with coercive pressures to use the relevant devices. The military context might be especially relevant in this regard, given the expectation that members of the armed forces should obey orders from their superiors (Tennison and Moreno, 2012; Ienca et al., 2018; Erler and Müller, forthcoming). Coercive pressures to use invasive BCIs seem less relevant to the civilian sector, at least for the foreseeable future.

What about the coercion issue when divorced from safety concerns? In a professional context, it might especially apply to non-invasive BCIs. An employer might for instance mandate that her employees wear EEG headsets to monitor, and ultimately enhance, their focus at work. This might strike some as intrinsically problematic. However, unlike coercion to use invasive BCIs, which could be viewed as infringing on people's right to bodily integrity, it seems less obvious that coercion to use non-invasive devices must be problematic *per se*. Indeed, one might argue that it is no different from existing and widely accepted forms of coercion aimed at enhancing work performance, such as mandatory employee training programs (Erler, 2020).

That being said, one might plausibly adduce *distinct* considerations in support of the view that there is something uniquely problematic about coercion to use (non-invasive) BCIs: say, that it presents a threat to the users' privacy or cognitive liberty. We will consider these separately, in the next section.

### *14.3.3 Privacy and cognitive liberty*

To some extent, the issue of privacy in this context is simply an extension of existing concerns raised by current practices, such as the use of the internet, AI assistants, various "smart" systems, and wearables (Müller, 2020). Yet even though future AI advisors may not raise any fundamentally new concerns in this regard when they do not involve the collection of *brain* data, they are still likely to further intensify existing ones. Tech companies like Apple and Amazon are thus known to be using human contractors to listen to users' recorded conversations with their digital assistants Siri and Alexa (Gartenberg, 2019). Given that many people already feel uneasy about such practices, they will likely have even greater objections if they were applied to the interactions about highly personal matters they might have with their moral or health AI advisor. The risk of data theft would be another concern. This highlights the need for sound policies on data privacy and protection, as well as the promotion of informed consent among users. The latter goal, of course, is a particularly challenging one in an ever-more complex digital world: recent surveys thus find that more than 90% of internet users agree to terms of service they have not read (Guynn, 2020).

A relevant question is whether the sheer collection of brain data, via devices like BCIs, makes a fundamental difference to the privacy issue. What would matter is to extract some meaning from this data, e.g. *decoding* the cognitive content, e.g. down to specific thoughts or attitudes. Such extraction will often be probabilistic and depend on other data sources, including previously observed behavior. However, given the special status usually attributed to the privacy of thought, people could reasonably feel reluctant to use such devices for IA if they knew they were *capable* of decoding content, even if they had been assured that no such intrusive data collection and decoding would occur. To this, we should add the risk that BCIs might get hacked by malicious actors. While having one's private thoughts exposed is already a possibility today, the development of true brain-reading technology would still represent the fall of the last bastion of mental privacy, calling for careful regulation.

Experts diverge about the likelihood that such a development might occur in the foreseeable future (compare for instance Ienca et al., 2018, with Wexler, 2019, and Gilbert et al., 2019). For now, however, we can already ask whether requiring employees or students to consistently use a device that monitored their attention levels would be an infringement on their right to mental privacy. It is not clear to us that the answer must be positive, insofar as attention levels are not typically considered mental states of the kind deserving strict protection from scrutiny in such contexts:

it is, for instance, not improper for a teacher to observe her students' behavior in class and call out those who appear to be distracted.

Perhaps a stronger reason to object to such monitoring is its potential negative *psychological* impact on users, who might feel that they are operating under oppressive surveillance, or being treated in a patronizing manner. Furthermore, inappropriate lessons could also be drawn from the data collected – an issue we address in Section 14.3.5. Finally, even if it is not assumed to disrespect mental privacy, it may be that attention monitoring for IA still violates another right, or as some would say in this context, "neuro-right" (Yuste et al., 2017): namely cognitive liberty, or freedom from unwanted interference with one's neural processes (Bublitz, 2013). Whether or not it does so will depend on the specific scope of that right. For instance, should we think that people have a right to occasionally allow their minds to wander at work, say as a natural way of alleviating boredom? If so, pressures toward attention augmentation could infringe on that right. Seeking to foster constant laser-like focus among students and employees could also prove counterproductive, insofar as mind-wandering appears to be conducive to creative thinking (Fox and Beaty, 2019). However, this is a consequentialist objection to the practice, rather than a rights-based one.

Other violations of cognitive liberty might involve hackers taking control of the relevant AI devices. Such "brain hacking" (Ienca and Haselager, 2016) could be done for the purpose of stealing sensitive brain data but also to seize control of an external device receiving commands from a BCI, or of the input to the user's brain, in the case of bi-directional BCIs. While the first of these three possibilities would again threaten mental privacy, the third one would conflict with cognitive liberty and "mental integrity" (Ienca and Andorno, 2017; Lavazza, 2018). Both civilians and military personnel could be the targets of such attacks, highlighting the importance of striving to incorporate adequate protections into the design of such devices.

### *14.3.4 Authenticity and mental atrophy*

Concerns about "authenticity" are recurrent in discussions of the ethics of enhancement, generally (The President's Council on Bioethics (US), 2003; Erler, 2014). We have already discussed the possibility that some AI advisors or BCIs might not deliver on their promises. Yet even assuming that such devices would allow for better decisions and improved performance, some might still object that they would offer a mere simulacrum of what they ought to be providing. This charge might particularly apply to forms of IA that did not count as full-fledged enhancements. For instance, one might contend that even if a moral AI advisor did provide us with sound ethical advice, it would nevertheless fail to authentically enhance our capacity for moral reasoning, since it would be delivering the end result of such reasoning "on a plate," circumventing the need to effortfully work things out for ourselves. In fact, one might fear that regularly outsourcing moral and prudential reasoning to AI advisors would cause our own capacity to engage in such reasoning to atrophy due to insufficient practice. IA would then entail a form of regression (Danaher, 2018), which is a version of the general "autopilot problem."

Several replies can be given to this argument. First, one might deny that relying on an AI advisor rather than exercising our own judgment, even on important matters, must be problematic. After all, life calls on us to make a large number of decisions, not all of which we may value *intrinsically*. Suppose for instance that Theodore wants to successfully invest his savings for retirement, without relishing the prospect of learning all the tricks of the art of investing. It is not plausible to think that he would deserve criticism for delegating most of his investment decisions to a trusted financial advisor, human or artificial. Theodore could rationally decide to devote his time and energy to other pursuits he considered more rewarding, and accept the resulting underdevelopment of his investment skills.

This reply does go some way toward answering the concern about authenticity and mental atrophy, yet it does not seem equally persuasive in relation to *all* forms of AI outsourcing. The moral domain stands out again here. Someone who systematically deferred to a moral AI advisor when making weighty ethical choices, and could not justify such a choice by themselves, but only repeat the rationale provided by the device, would arguably have failed to develop an important aspect of themselves as a human person.

To avoid such problem cases, one might instead try and respond to the authenticity concern by invoking the extended mind thesis (EMT): namely the claim that external artifacts can become part of an individual's mind or cognitive system if the right conditions are met. As stated by one of the EMT's original proponents, Andy Clark, such conditions include: 1) that the resource in question be reliably available and typically invoked; 2) that any information retrieved from it be more or less automatically endorsed; and 3) that the relevant information be easily accessible as and when required (Clark, 2010). Suppose now that Theodore's moral AI advisor is consistently available to him (via his smartphone or otherwise) wherever he goes and regularly gives him advice which he always accepts without hesitation. Based on the EMT, one might then contend that when the AI advisor engages in ethical reasoning, Theodore does so authentically too, since the device has become an extension of his mind. Clearly, the main vulnerability of this reply is that it stands or falls with the EMT, which is a controversial view (see Coin and Dubljevic, 2021).

Perhaps a stronger response to the concern is that while *some* uses of AI advisors for IA might indeed problematically supersede significant human activities like ethical reasoning, many need not do so. For instance, the "Socratic" moral advisor proposed by Lara and Deckers arguably would not, since it would be designed to prod the user into working out sounder ethical opinions by themselves, rather than delivering "ready-made" advice. The same applies to devices that would simply alert users to ethically risky physiological states, or help them track their progress toward their philanthropic goals. Finally, even devices that did recommend specific solutions to ethical dilemmas need not be blindly obeyed. Users could thus choose to override their advice based on their own reasoning, or only decide to follow it after having carefully considered the justification provided by the device. This could arguably enrich a person's moral thinking, rather than substitute for it. Overall, it seems that the considerations of autonomy are the same for human and IA advisory systems.

### 14.3.5 Fairness

If applications of AI for IA deliver on their promise, they risk exacerbating the existing "digital divide" (Ragnedda and Muschert, 2013), both at a local and global level. Those who cannot afford the relevant devices, or who are not able to access them via their school or employer, might unfairly find themselves at a disadvantage compared to those who can, in various important life domains from career to health. How to tackle this issue can be viewed as one aspect of the larger, momentous challenge of promoting equitable access to beneficial technologies, a challenge to which there is no simple solution. Beyond inequality of access, differential benefits might also result from differences in digital literacy, including "BCI literacy" (Cinel et al., 2019). While training programs designed to boost people's proficiency at using the relevant devices can help overcome those differences to some extent, other solutions might be required in cases where such differences are not grounded in unequal learning opportunities – but rather, say, in individual differences in the brain activity to be captured by BCI devices.

If used judiciously, the use of BCIs to monitor and augment attention at school and work could have benefits for users beyond any enhancing impact on performance. For instance, it could make it easier to identify students with conditions like attention deficit hyperactivity disorder (ADHD) or gifted students who find their classes insufficiently challenging, and to compare the degree of student engagement achieved by different teaching styles. Constructive adjustments could then be made. Nevertheless, we might also want to warn against inappropriate action being taken in response to the information thus collected. Students with ADHD, or those who get bored by a course that is insufficiently engaging or too easy for them, should not get unfairly penalized for their low attention scores.

Similar remarks apply to workplace applications. McCall and Wexler note that "some individuals may have a higher performance level than others even when in a distracted state" (2020, p. 14). It is plausible to think that professional rewards should be tied to a worker's absolute performance level, rather than to their overall level of attention and wakefulness at work. That said, it also seems that a company could reasonably use the relevant data to try and optimize employee performance, by appropriately adjusting working conditions. This could include making space for naps: a 2016 report by the RAND Corporation thus estimated that the United States lost an equivalent of around 1.23 million working days each year from sleep-deprived workers (Hafner et al., 2016).

An additional fairness-related concern is the issue of algorithmic bias. Even without any intention on the part of their designers, AI advisors could exhibit biases in their recommendations that unfairly disadvantaged certain social groups, whether as a result of the procedure used to issue those recommendations, or of the data on which they would be relying. A health AI advisor, for instance, might not provide equally reliable input to users from ethnic minorities, if the data used to train its algorithms did not feature enough members of those underrepresented groups (Kaushal et al., 2020). This is therefore an issue to be monitored, as part of the broader phenomenon of AI bias. Potential solutions include ensuring that the relevant algorithms are trained on data sets derived from diverse populations. While much more research remains to be done on this topic, it is worth noting that the presence of bias in an AI system does not automatically imply the preferability of relying solely on human decision-making, since biases among people can be even greater (Ledford, 2019).

## 14.4 Conclusion

This overview of the paths toward mustering the power of AI for IA, including human enhancement, suggests that they hold real promise, while also raising several ethical concerns. Some of these concerns apply across the board, while others differ based on the type of AI application being considered, such as whether it constitutes enhancement or "sheer" IA. The former concerns include the risk that the relevant devices might not perform as expected, threats to data privacy, and issues of equitable access and AI bias. As for the latter concerns, forms of IA that are not full-fledged enhancements seem more vulnerable to objections relating to inauthenticity and mental atrophy. Those that do count as enhancements, by contrast, might be more likely to present threats to cognitive liberty, insofar as they truly involve altering a person's cognitive functioning and elicit safety concerns, as such alterations of functioning might necessitate more invasive interventions. This last point, however, might depend on the position one takes on the extended mind thesis.

As new narrow forms of superintelligent AI get developed, the resulting risk of human obsolescence, both in the professional domain and in activities often considered significant aspects of existence, such as moral deliberation, will likely increase the rationale for pursuing a genuine fusion between humans and machines of the kind advocated by some transhumanists, rather than continuing to treat AI as a useful tool. That said, given the sizable technical challenges to such a prospect, more mundane applications of AI for IA, and their associated ethical conundrums, will remain with us for some time.